\documentclass[]{interact}

\usepackage{epstopdf}
\usepackage[caption=false]{subfig}

\usepackage[numbers,sort&compress]{natbib}
\bibpunct[, ]{[}{]}{,}{n}{,}{,}
\makeatletter
\def\NAT@def@citea{\def\@citea{\NAT@separator}}
\makeatother

\theoremstyle{plain}

\theoremstyle{definition}

\theoremstyle{remark}

\begin{document}


\title{Icosahedral order in liquid and glassy phases of cyclohexane}

\author{
\name{Tomoko Mizuguchi\thanks{CONTACT T. Mizuguchi. Email: mizuguti@kit.ac.jp} and Soichi Tatsumi and Susumu Fujiwara}
\affil{Faculty of Materials Science and Engineering, Kyoto Institute of Technology, Matsugasaki, Sakyo-ku, Kyoto 606-8585, Japan}
}

\maketitle

\begin{abstract}
We performed all-atom molecular dynamics simulations for bulk cyclohexane and analysed the short- and medium-range structures in supercooled and glassy states by using the Voronoi tessellation technique. From the analyses of both the potential energy of the system and the radial distribution function of molecules, cyclohexane was found to be vitrified as the temperature decreased. Furthermore, the icosahedral-like structures are dominant at all temperatures and grow in a supercooled liquid, whereas the face-centred cubic structures do not grow when the temperature decreases. It was also ascertained that the icosahedral-like structure is more dominant than the full-icosahedral one. 
The network of the distorted icosahedron spreads throughout the system at low temperatures.
Our simulation demonstrates the stability of the icosahedral structure even in a non-spherical molecule such as cyclohexane.
\end{abstract}


\section{Introduction}

In the glass transition process, the motion of particles slows down drastically as the temperature is decreased.
This is accompanied by the divergence of relaxation time or viscosity in a small range of temperatures.
Although many experimental, theoretical, and computational techniques have been employed in an attempt to elucidate the origin of glass transition~\cite{Angell2000,Berthier2011,Kob1999,Tarjus2005,Cavagna2009}, none of them has succeeded in explaining all abnormal behaviours related to the glass transition.

To avoid crystallization, many experiments on glass transition have been conducted on molecules with complex structures, such as polymers~\cite{Cangialosi2014,Roth2016}. 
In recent years, advances in experimental techniques have made it possible to vitrify even simple molecules such as propene, benzene, and cyclohexane~\cite{Tatsumi2012,Xia2006,Tatsumi2015,Yamamuro2003}. 
Interestingly, cyclohexane, which exhibits the glass transition by confinement in nanopores with a diameter of 1.9 to 2.9 nm, demonstrates a heat capacity anomaly caused by the first-order phase transition at a higher temperature than the glass transition point~\cite{Tatsumi2015}.
This means that certain structural changes should occur in supercooled liquid cyclohexane, but the changes are small and thus difficult to capture experimentally.

Additionally, many computer simulations of supercooled liquids and glasses have been done with simple isotropic models.
With recent developments in computer technology, it has become possible to realistically model and simulate realistic molecules at the atomic level.
We use an all-atom model of cyclohexane as a realistic model of a molecular liquid for analysis in supercooled and glassy systems.

\begin{figure}
\includegraphics[width=10cm]{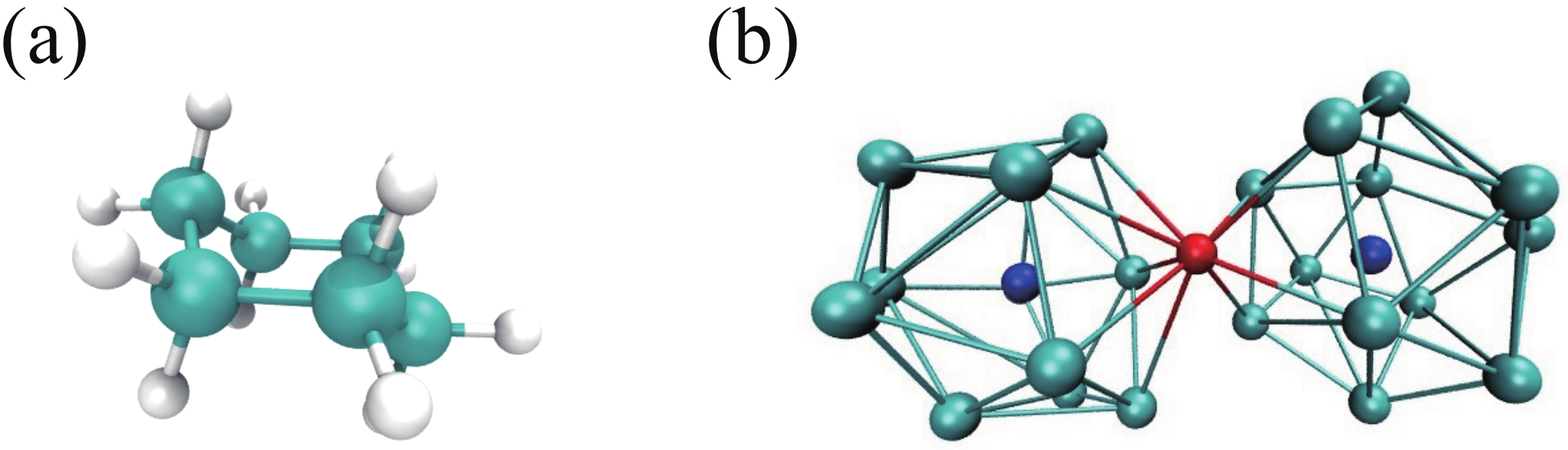}
\caption{(colour online) (a) A cyclohexane molecule represented by balls and sticks. Cyan and white spheres represent carbon and hydrogen atoms, respectively. (b) An example of the connection of the molecular clusters. The sphere represents a cyclohexane molecule. Cyan, blue, and red spheres represent nearest-neighbor shells, central molecules, and shared molecules, respectively. The bond between spheres is not a real bond and is simply shown for clarity.}
\label{fig1}
\end{figure}

Cyclohexane is a small molecule, with the chemical formula is C$_6$H$_{12}$ (see Fig.~\ref{fig1}(a)), and it mainly interacts via the van der Waals (vdW) force with like molecules.
The melt is transformed at 280 K into a plastic phase, which has a cubic cell, as well as at 186 K into a crystalline phase, which has a monoclinic cell~\cite{Kahn1973}.
However, crystallization can be prevented by confinement within a nano-sized pore, as described above.
The focus of this work is the short- and medium-range structures in supercooled cyclohexane.
We examine the change in the local structures using molecular dynamics (MD) simulations and show that icosahedral structures grow in the supercooled state.

In Section~\ref{sec:method}, we explain the methods used in the MD simulations as well as the structural analysis. 
In Section~\ref{sec:result}, we analyse the short- and medium-range structure of supercooled and glassy cyclohexane using Voronoi tessellation.  
In Section~\ref{sec:summary} we close with concluding remarks.

\section{\label{sec:method}Methods}
\subsection{Simulation procedure}
We carried out all-atom MD simulations of bulk cyclohexane using the CHARMM Drude force field developed by Vorobyov {\it et al.}~\cite{Vorobyov2007} in the NAMD2.10 program package~\cite{namd}.
Periodic boundary conditions were applied to the MD unit cell which contains 1372 cyclohexane molecules.
The vdW interaction is represented by the Lennard-Jones potential which was truncated by applying a switching function~\cite{switch} with a range of 10--12 \AA.
The long-range electrostatic interaction was calculated using the smooth particle-mesh Ewald method~\cite{spme}.
All simulations were performed in the isothermal-isobaric ensemble with a time step of 1 fs.
The pressure was maintained at 1 atm by the Langevin piston Nos{\'e}-Hoover method~\cite{npt1,npt2} with barostat oscillation and damping time constants of 100 and 50 fs, respectively.
The temperature was controlled with the Langevin dynamics at a damping coefficient of 1.0 ps$^{-1}$.
The SHAKE algorithm~\cite{shake} was used to fix the bond lengths involving the hydrogen atoms.

First, we prepared liquid cyclohexane at 360 K and quenched the system to 10 K at a cooling rate of $10^{11}$ K/s.
Starting from the configurations dumped during the cooling process, we conducted 5-ns runs for structural relaxation and subsequent 1-ns production runs at each temperature.

\subsection{Structural analysis}
The intermolecular structures were analysed in terms of Voronoi polyhedra.
First, we calculated the centre of mass for each molecule from the MD snapshots.
Next, the Voronoi diagram was computed for the centre-of-mass coordinates using a Voro++ library~\cite{Rycroft2009}. 
Finally, for each Voronoi polyhedron, each face with an area less than 1\% of the total surface area was removed in order to avoid overcounting~\cite{Brostow1998}.
We used the Voronoi index $\left< {n_3,n_4,n_5,n_6} \right>$ to identify the type of each polyhedron, where $n_i$ is the number of faces with $i$ vertices.
To characterise the medium-range order structures, we define the cluster as the nearest neighbour shell obtained by Voronoi tessellation.
The network is defined as the connection of the clusters: two clusters are considered to be connected if they share at least one molecule~\cite{Ward2013}.
An example of the connection of the clusters is shown in Fig.~\ref{fig1}(b).

\section{\label{sec:result}Results and Discussion}
\begin{figure}
\includegraphics[width=8cm]{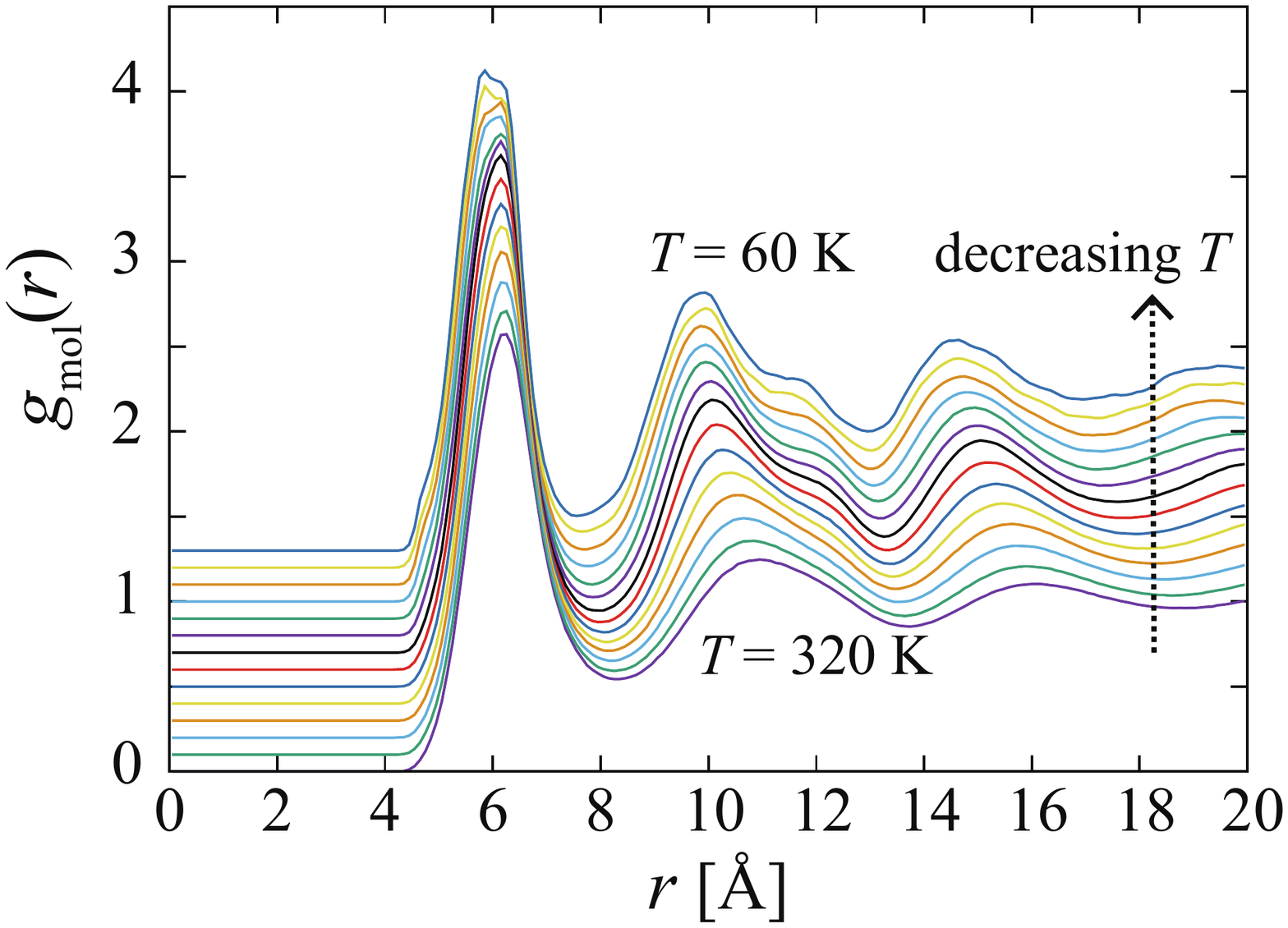}
\caption{(colour online) Radial distribution functions between the centre of mass of cyclohexane molecules at various temperatures ranging from 60 K to 320 K with 20 K step. Each line is shifted 0.1 units with respect to the next temperature for readability.}
\label{fig2}
\end{figure}

First, we calculated the radial distribution functions for the centre of mass of the molecules $g_{\rm mol}$($r$) at each temperature.
Figure~\ref{fig2} shows $g_{\rm mol}$($r$) at various temperatures.
When the temperature was lowered, the peak positions of $g_{\rm mol}$($r$) shifted to smaller $r$ values, reflecting the increase in the density of the system. 
The second peak of $g_{\rm mol}$($r$) has a small sub-peak around $r$ = 12 \AA\ at lower temperatures.
This corresponds to the well-known observation that the second peak of $g_{\rm mol}$($r$) splits in a glassy state~\cite{Kob1995,Bailey2004,Chen2017}.
The split second peak of $g_{mol}$($r$) is considered to be related to the icosahedral local order, which can be seen in disordered systems composed of isotropic particles such as Lennard-Jones systems~\cite{Kob1995}, bulk metallic glass~\cite{Bailey2004}, and random hard-sphere packing~\cite{Clarke1993}.
In our systems also, icosahedral local order was observed, as described later.

\begin{figure}
\includegraphics[width=7cm]{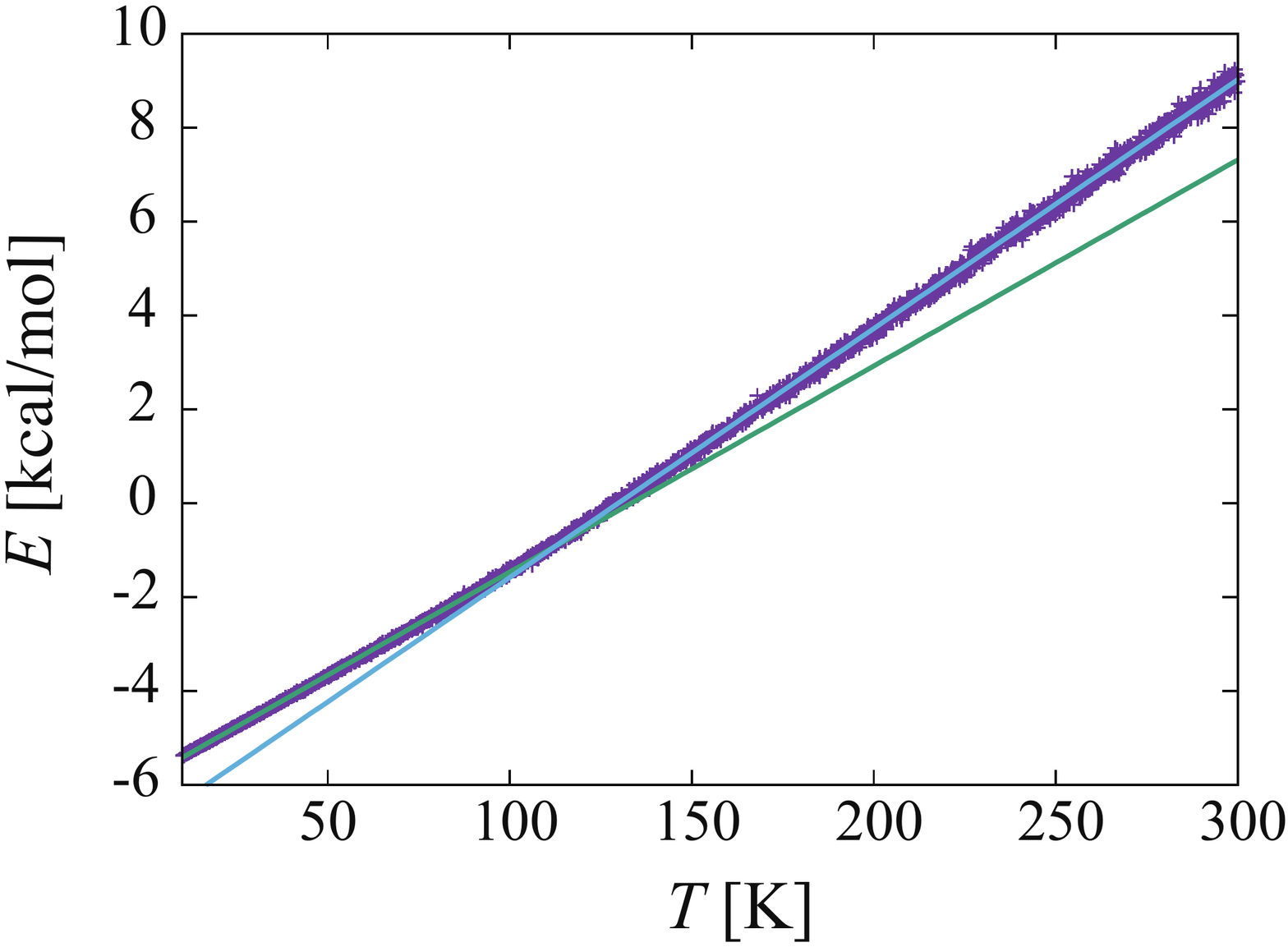}
\caption{Temperature dependence of potential energy per molecule. Solid lines are fitting functions.}
\label{fig3}
\end{figure}

Potential energy per molecule is shown in Fig.~\ref{fig3} as a function of temperature. 
The slope changes around 110 K but there is no jump and thus it does not crystallise and instead forms a glass at low temperatures.
We can evaluate $T_{\rm g}$ from the figure as the point where the slope changes~\cite{Bailey2004}.
$T_{\rm g}$ is determined by dividing the curve into two pieces and fitting a straight line to each; the intersection of the  two lines is defined as $T_{\rm g}$.
From Fig~\ref{fig3}, $T_{\rm g}$ was estimated to be approximately 112 K.
This value is in agreement with the experimental value obtained from heat capacity measurements of confinement cyclohexane, in which $T_{\rm g}$ was 80--100 K~\cite{Tatsumi2015}.
Note that the higher $T_{\rm g}$ in simulations than experiments can be attributed to the faster cooling rate in the simulations than that in experiments. 

\begin{figure*}
\includegraphics[width=15cm]{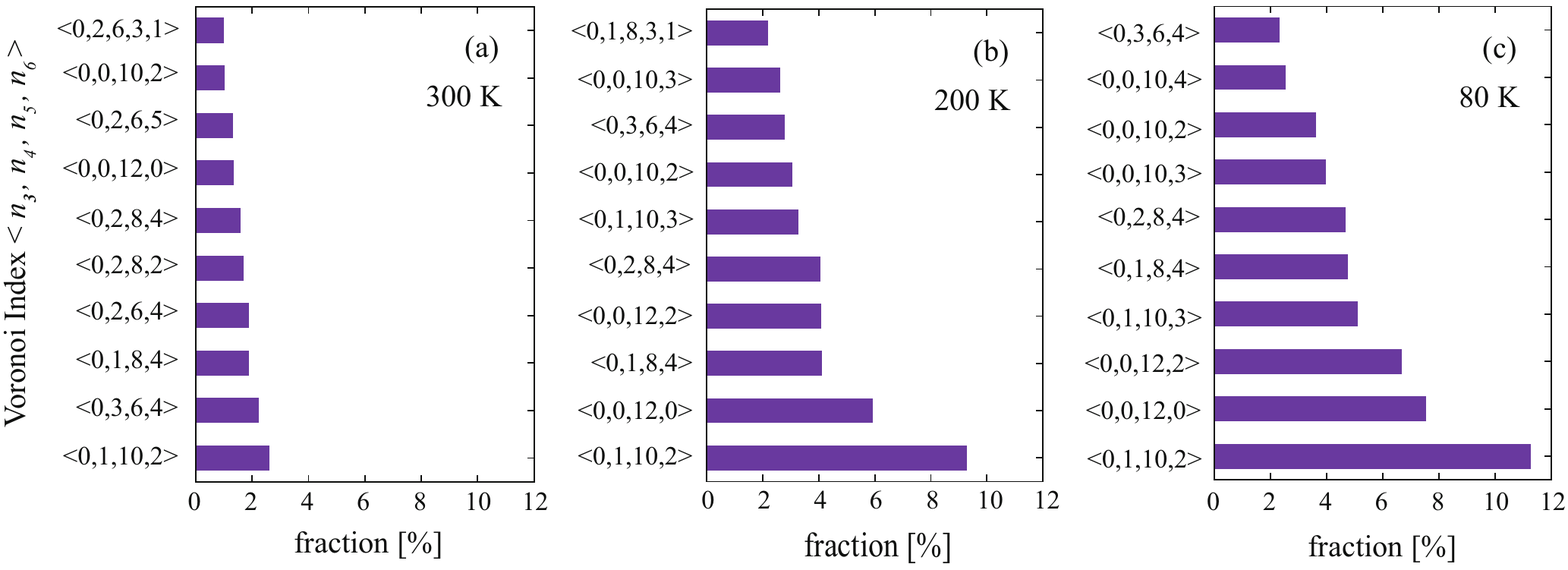}
\caption{Fractions of the ten most populous Voronoi polyhedra at (a) 300 K, (b) 200 K, and (c) 80 K.}
\label{fig4}
\end{figure*}

To examine the local intermolecular structures, we performed Voronoi tessellation for the centre-of-mass coordinates.
Figure~\ref{fig4} shows the fractions of the ten most populous types of polyhedra with the Voronoi index $\left< n_3,n_4,n_5,n_6 \right>$ at $T$= 300 K, 200 K, and 80 K. 
The $\left<0,1,10,2\right>$ type of polyhedron is the most populous at all three temperatures, although its fraction at $T$= 300 K is not much different from that of other polyhedra.
It is an icosahedral-like cluster and is frequently observed in metallic glasses~\cite{mauro2011,hirata2013,Zhang2015,Yang2017,Shimono2015,Fukunaga2007,Ward2013,Ganesh2008}.
The full icosahedron $\left<0,0,12,0\right>$ was also observed; however, the fraction was smaller than that of $\left<0,1,10,2\right>$.
This implies that an icosahedral-like cluster is preferred as a local structure rather than a full icosahedron in the supercooled and glassy state of cyclohexane.

\begin{figure*}
\includegraphics[width=14cm]{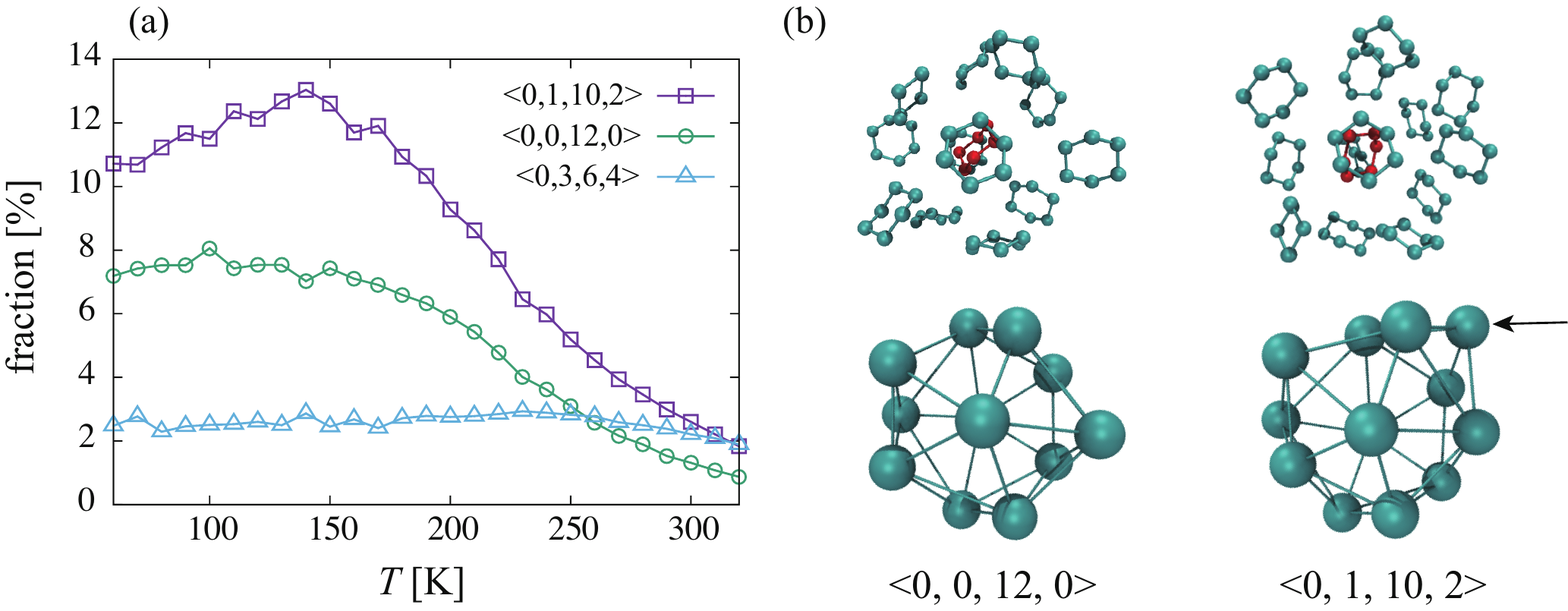}
\caption{(colour online) (a) Temperature dependence of the fractions of $\left<0,1,10,2\right>$, $\left<0,0,12,0\right>$ and $\left<0,3,6,4\right>$ polyhedra. (b) Structures of $\left<0,0,12,0\right>$ and $\left<0,1,10,2\right>$ polyhedra. In the upper panels, only carbon atoms are displayed and the centre molecule is coloured red. In the lower panels, a cyclohexane molecule is represented by balls and the bond between balls is shown only for clarity (not real bonds).}
\label{fig5}
\end{figure*}

The plastic phase of cyclohexane has a face-centred cubic (fcc) lattice, whose Voronoi polyhedron is $\left<0,12,0,0\right>$.
However, this polyhedron, as well as other crystal polyhedra such as $\left<0,6,0,8\right>$ in a body-centred cubic lattice are not seen at all temperatures.
Conversely, we do see $\left<0,3,6,4\right>$ polyhedra at all temperatures, which is considered to be an fcc-like cluster~\cite{Cape1981,Shimono2015,Jiang2016}.
However, the fraction of $\left<0,3,6,4\right>$ polyhedra is small and almost constant at 2--3 \% at all temperatures, and importantly it does not grow when temperature decreases.
The temperature dependence of the fractions of $\left<0,1,10,2\right>$, $\left<0,0,12,0\right>$ and $\left<0,3,6,4\right>$ polyhedra is shown in Figure~\ref{fig5}(a).
As seen in the figure, the icosahedral structures $\left<0,1,10,2\right>$ and $\left<0,0,12,0\right>$ grow in a supercooled liquid.
We show the simplified diagrams of $\left<0,0,12,0\right>$ and $\left<0,1,10,2\right>$ clusters in Fig.~\ref{fig5}(b).
The index $\left<0,0,12,0\right>$ represents a full icosahedron but the $\left<0,0,12,0\right>$ cluster in our system is not necessarily a perfect icosahedron.
Hirata {\it et al.}, reported that icosahedral clusters observed in Zr$_{80}$Pt$_{20}$ metallic glass are not ideal icosahedra but are distorted due to the geometric frustration of icosahedral order~\cite{hirata2013}.
Additionally, in our system, the $\left<0,0,12,0\right>$ cluster is distorted, and the $\left<0,1,10,2\right>$ cluster has an icosahedral-like symmetry but has one extra molecule (indicated by an arrow in Fig.~\ref{fig5}(b)).

\begin{figure}
\includegraphics[width=8cm]{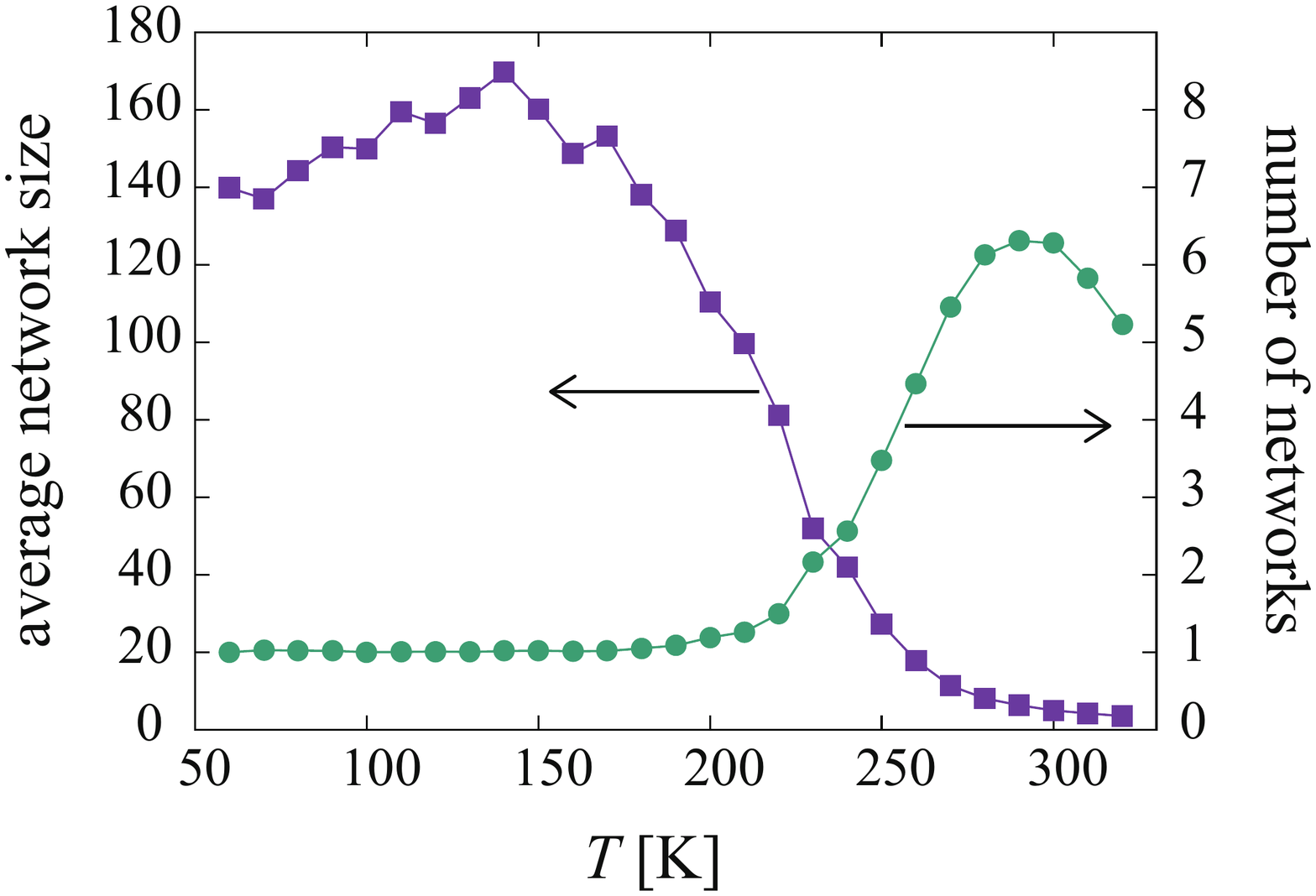}
\caption{(colour online) Average network size (purple square, left scale) and the number of networks (green circle, right scale) of $\left<0,1,10,2\right>$ clusters as a function of temperature.}
\label{fig6}
\end{figure}

\begin{figure}
\includegraphics[width=15cm]{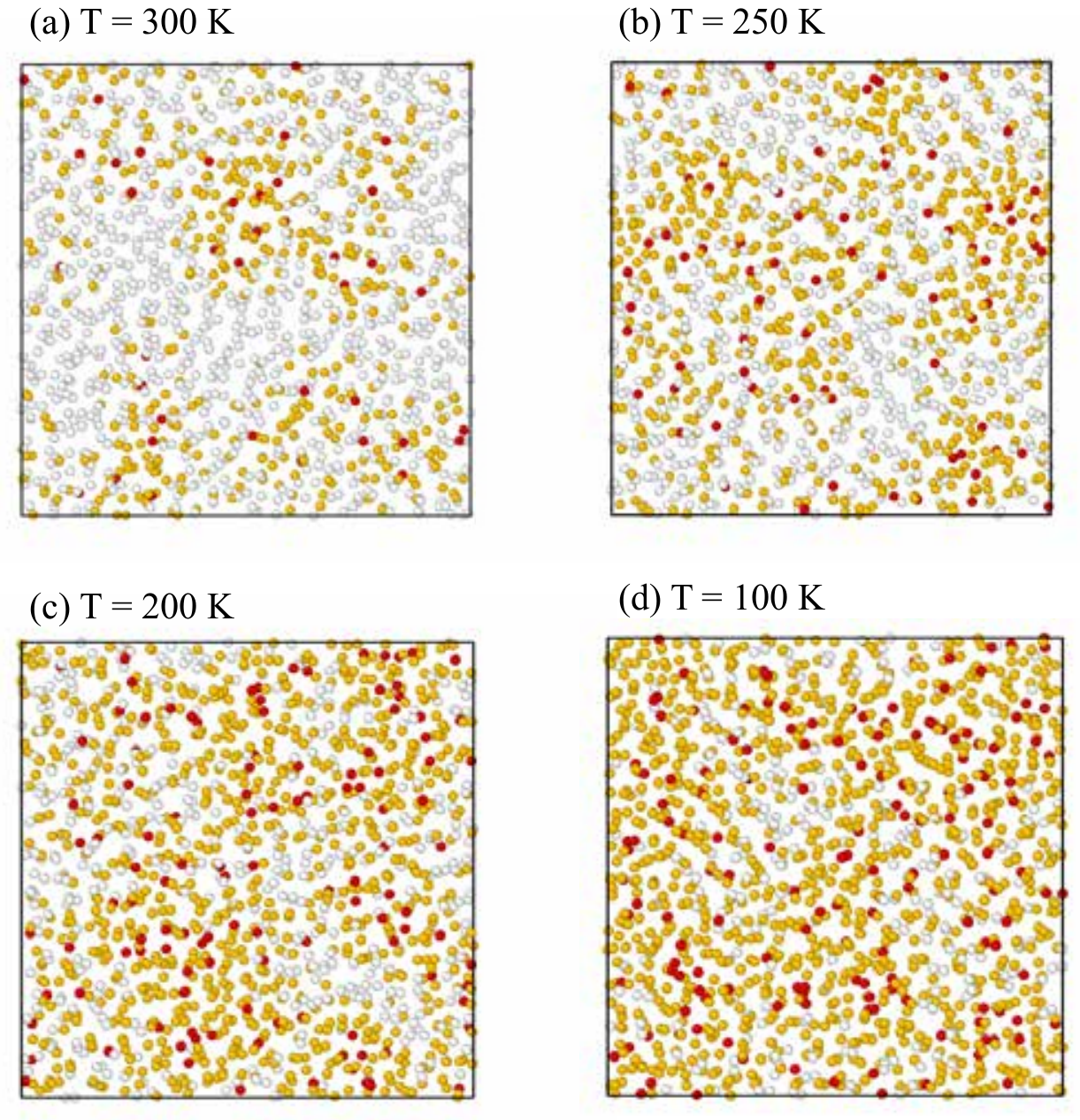}
\caption{(colour online) Spatial distributions of $\left<0,1,10,2\right>$ clusters at 300, 250, 200, and 100 K are shown as a projection on a two-dimensional plane. The centre-of-mass coordinate of a cyclohexane molecule is represented by a sphere. Red spheres represent the centre molecules in $\left<0,1,10,2\right>$ clusters, and orange spheres are molecules that are composed of $\left<0,1,10,2\right>$ clusters excluding the centre molecules. Others are shown in light grey.}
\label{fig7}
\end{figure}

To investigate the medium-range order, we calculated the average network size and the number of networks of $\left<0,1,10,2\right>$ clusters as shown in Figure~\ref{fig6}.
The temperature dependence of the average network size is similar to that of the fraction of the $\left<0,1,10,2\right>$ polyhedron (Fig.~\ref{fig5}(a)).
The number of networks becomes approximately 1 below 200 K, meaning that the network spreads throughout the system.
The spatial distribution of $\left<0,1,10,2\right>$ clusters is shown in Figure~\ref{fig7} as a projection on a two-dimensional plane.
A cyclohexane molecule is represented by a sphere that is located at the centre-of-mass coordinate of the molecule.
Red spheres represent the centre molecules in $\left<0,1,10,2\right>$ clusters, and orange spheres are molecules that are composed of $\left<0,1,10,2\right>$ clusters excluding the centre molecule.
At 300 K, several networks exist separately.
At 250 K, the size of each network becomes larger but some of them are still separate.
At 200 K, the networks grow and merge into nearly a single network.
At 100 K, the density of network increases and the $\left<0,1,10,2\right>$ cluster spreads throughout the system.

\section{\label{sec:summary}Summary and Conclusions}
We performed all-atom MD simulations of the liquid and glassy phases of cyclohexane as a model of a molecular liquid.
The glassy state was achieved by quenching from the melt and confirmed by the energy change and radial distribution function.
The local structure was analysed using Voronoi tessellation and icosahedral structures were found in the supercooled and glassy state.
The dominant structure is an icosahedral-like $\left<0,1,10,2\right>$ structure rather than a full-icosahedral $\left<0,0,12,0\right>$ structure.
The $\left<0,1,10,2\right>$ polyhedral network grows in three dimensions as the temperature decreases and spreads throughout the system.
The icosahedral structure is considered to be important in supercooled liquids and glasses because it is highly close-packed~\cite{Frank1958} and causes geometrical frustration owing to inconsistencies in the translational symmetry.
In fact, it has been observed in many metallic glasses.
In this study, we found the icosahedral order in supercooled cyclohexane even though the shape of cyclohexane is not necessarily spherical.
This observation reinforces the idea that the icosahedral order plays a role in glass formation.
The relation between the experimentally observed transition in confined cyclohexane and its local structures are currently under investigation.
Furthermore, an all-atom model can incorporate even intramolecular structural changes, but it has a disadvantage for investigating slow dynamics near the glass transition point.
Long-time simulations using a coarse-grained model are for future research.

\section*{Acknowledgements}
This research used computational resources of the Supercomputer Center, the Institute for Solid State Physics, the University of Tokyo, and of the MEXT Joint Usage / Research Center "Center for Mathematical Modeling and Applications", Meiji University, Meiji Institute for Advanced Study of Mathematical Sciences (MIMS). 

\section*{Funding}
This work was partially supported by the Japan Society for the Promotion of Science KAKENHI grant no. JP19K05209.

\end{document}